# Fluency Without Evidence: Constraint-First Design and the Limits of Self-Report in AI-Assisted Learning

Fatima T. Zahra[1], Wei Wang[1], Frances Harper[1], Jiangen He[1]

[1]University of Tennessee at Knoxville, Knoxville, Tennessee, United States

**Abstract**

A generative AI teaching partner should support reasoning over supplying conclusions; however, this has not been tested against learning in an authentic course. Drawing on design-based research, we specify the position as a conjecture map and report a first design cycle in two graduate-level research methods courses. Students used an AI teaching partner employing a constraint-first sequence requiring them to state and justify positions before receiving questions. Pre- and post-measures of AI literacy, critical thinking, and metacognitive awareness were collected alongside interaction records. AI literacy increased, concentrating in understanding AI, whereas critical thinking, awareness, and knowledge did not change. Since changes were limited to self-report measures, they may reflect growth in confidence instead of capacity. Interaction records, meanwhile, showed brief exchanges, uneven enactment of the constraint-first sequence, and missing records. These findings show why AI-supported learning requires interaction records to provide a more defensible basis for AI-supported designs than self-reports.

## 1 Introduction

Research on generative AI in education has over the past several years settled on a fairly consistent normative position, which holds that these systems ought to function as thinking partners rather than as engines for producing answers. Klein et al. gave this position its fullest theoretical statement in arguing that the frictionlessness of generative retrieval creates a paradox where tools capable of extending intellect may equally erode it, and in proposing a Dual-ZPD framework that pairs Vygotskian proximal development with what they termed a zone of proximal motivation (Klein et al., 2026). Mao et al. pursued the same problem through design, observing that Socratic instruction depends on response strategies which many learners do not possess and developing scaffolding cards intended to supply them (Mao et al., 2026). Both draw on a body of evidence concerning the metacognitive demands that generative systems place on their users (Tankelevitch et al., 2024) and the cognitive offloading which follows when those demands go unmet (Lee et al., 2025). The concern is not confined to research communities. MIT's Ad Hoc Committee on AI Use in Teaching, Learning, and Research Training reported in 2026 that overreliance on chatbots may diminish critical thinking, weaken memory, erode confidence, and undermine mastery, while describing the available evidence for these effects as early signals and calling for a deeper scientific understanding of how AI affects learning (MIT Ad Hoc Committee on AI Use in Teaching, Learning, and Research Training, 2026).

However, despite the elaboration this position has received, what the literature has not produced is a test of it against measured learning or higher-order thinking, including metacognition and critical thinking, outcomes in an authentic course setting. Recent work at CHI has examined instructor experience with campus-wide AI tutoring platforms (Ko et al., 2026), teacher authoring of educational simulations (Kaputa et al., 2026), teachers' expectations concerning generative AI and educational inequality (Xiao et al., 2026), and short-term learner interaction with designed scaffolds (Mao et al., 2026); yet across this body of work, learner outcomes are theorized and designed for, but not directly measured, which leaves the underlying proposition, namely that positioning AI as a partner rather than as an authority improves reasoning, in the position of a conjecture that has not been examined.

It might be charged that such a central proposition ought not to rest on so little, and this paper accordingly approaches it as a conjecture. Working within design-based research, we specify the proposition as a conjecture map (Sandoval et al., 2014) in which three design features are held to shift students away from confirmation-seeking and toward reasoning, and where that shift is in turn held to produce gains in higher-order thinking and AI literacy. The first we call constraint-first: students have to state and justify their own position before the model responds. The second restricts the partner to questions. The third frames the system for students as a statistical model, not an authority. Mediating the relationship is what we term epistemic positioning, by which we mean the stance a learner adopts toward AI as a source of knowledge, and which may range from treating the system as a validator of work already completed to engaging it as a co-thinker whose claims are weighed against evidence (Hoher & Pintrich, 1997; Kuhn, 1999). We use students' dialogues with the AI teaching partner as interaction records to examine how the design was enacted. *We ask whether this conjecture holds when the design is implemented in authentic coursework, and what the interaction record reveals about how it was enacted.* This study asks:

1. What change is observed in students' higher-order thinking and AI literacy over a semester with a constraint-first teaching partner?

2. What does the interaction record show about whether the design was enacted as intended?
3. Taken together, what do the answers imply about how this literature measures these designs?

The study reported here is the first empirical cycle in which this conjecture has been tested. Two graduate research methods courses at a large public research university implemented the design across a semester, and we collected pre- and post-measures of AI literacy, critical thinking disposition, and metacognitive awareness alongside the complete record of students' interaction with the teaching partner. The paper thus contributes an empirical test of the co-thinker conjecture in authentic coursework, in which the interaction record rather than student self-report supplies the evidence about how students engaged; an account of what that record indicates concerning whether the constraint-first sequence operated as its designers intended; and an argument about measurement, prompted by the observation that everything which changed in this study was measured by self-report while everything which did not was measured by performance.

## 2 Related Work

### 2.1 From answer engine to thinking partner

The design ideal under examination here is well developed theoretically. Klein et al. situated the problem in the frictionlessness of generative retrieval, arguing that its very convenience produces a motivational deficit which conventional cognitive scaffolding is not equipped to address, and their Dual-ZPD framework accordingly identifies a productive learning zone in which both developmental and motivational conditions are satisfied (Klein et al., 2026). Mao et al. proceeded through design, noting that learners lacking effective response strategies produce superficial answers which undermine Socratic approaches, and supplying scaffolding cards to address that gap (Mao et al., 2026). The empirical motivation for both derives from studies of metacognitive demand (Tankelevitch et al., 2024) and of cognitive offloading among knowledge workers (Lee et al., 2025).

Our own account locates the problem differently, and the difference is a substantive one rather than a matter of terminology. Dual-ZPD is a motivational and individual framework which asks whether a learner is adequately supported and adequately motivated to undertake cognitive work instead of delegating it; epistemic positioning asks a relational question concerning what role the learner assigns the model and what that assignment does to the learner's own authority over the material. The framing follows Freire's account of banking pedagogy, where knowledge is deposited into learners rather than constructed with them (Freire, 2000) and draws on work in epistemic cognition concerning where learners locate the warrant for a claim (Hoher & Pintrich, 1997; Kuhn, 1999). The distinction carries practical consequences, since a highly motivated student may nonetheless treat a model as an authority whose output settles the question at hand, and a motivational account offers little purchase on the case.

### 2.2 Learner outcomes in the CHI education literature

Recent work at CHI on generative AI in education has concentrated on the people and institutions surrounding the tool rather than on what the tool does to learners. Ko et al. interviewed instructors deploying course-specific AI tutors through a campus-wide platform and identified tensions between institutional adoption and instructional practice (Ko et al., 2026). Kaputa et al. addressed teacher authoring of educational simulations through task-level abstraction, motivated by the absence of programming affordances that would allow teachers to verify simulations against the concepts they intend to convey (Kaputa et al., 2026). Xiao et al.

examined how teachers internationally anticipate the consequences of generative AI for educational inequality (Xiao et al., 2026). Mao et al. studied learners directly, though through an exploratory design study which did not measure outcomes (Mao et al., 2026), and Klein et al. contributed a framework unaccompanied by data (Klein et al., 2026).

Table 1 summarizes the pattern, which is that this literature studies perceptions, builds artifacts, and develops theory, while leaving largely unaddressed the question of whether learner thinking changes at all. Thus we see a body of design research whose central claims concern learning and whose evidence concerns almost everything except learning; and it is this gap the present study enters.

Table 1.

Recent CHI work on generative AI in education

| **Paper** | **Contribution type** | **Data** | **Learner outcomes measured** |
|---|---|---|---|
| Mao et al. | Design exploration | Exploratory study | No |
| Klein et al. | Theory framework | None | No |
| Ko et al. | Interview study | Instructors | No |
| Kaputa et al. | System and tool | Teacher authoring | No |
| Xiao et al. | Envisioning study | Teachers | No |

*Note.* Learner outcomes refer to measured change in learners, not to reported perceptions or design evaluation.

### 2.3 Design-based research as a bridge

Design-based research develops and refines theory and design together across iterative cycles conducted in authentic settings, treating context as a feature of the work, not as a confound to be controlled, and treating design revisions as themselves a form of evidence (Brown, 1992; Cobb et al., 2003; Collins, 1992). Because the tradition is well established within the learning sciences yet less commonly employed in Human-Computer Interaction (HCI), its commitments are stated explicitly here.

Conjecture mapping is the mechanism by which design-based research remains falsifiable over merely descriptive, since a conjecture map specifies the path running from design features through mediating processes to outcomes and thereby commits the researcher in advance to what would count as disconfirmation (Sandoval, 2014). The high-level conjecture in the present case is that students who position the AI as a co-thinker rather than as a validator or a generator engage in richer reasoning, and that they show stronger gains in higher-order thinking and AI literacy as a result. Stating it this way makes clear what evidence would challenge the claim. If positioning does not vary across students, or varies without bearing on outcome change, the conjecture is not supported. Two implications follow for the reading of this paper. The study is one cycle within a multi-cycle program, so its findings are interpreted as documented change and as grounds for refinement. And because early cycles exist in part to expose conjectures which do not hold, a conjecture finding no support at this stage is a result instead of a failure of the study. In this respect, the paper attempts to bring into contact two research traditions which ask different questions of the same designs.

# 3 Methods

## 3.1 Study context

The design was implemented in four graduate research methods courses at a large public research university in the southeastern United States during Spring 2026, and this paper reports the two for which data were returned. In the remaining two courses, departures from the data collection protocol meant that neither matched survey responses nor interaction records were collected, and those courses accordingly appear in no analysis reported here; we treat this as site-level attrition and consider its implications in Section 5.4.

Course A is a doctoral course in quantitative research methods, models, and measurement offered within a hospitality and tourism management program, while Course B is a graduate course in the foundations of educational data science, in which students work with R, reproducible workflows, and applied data analysis. Both are core methods courses within their respective programs, and both are housed in a college of education, health, and human sciences. A total of 26 enrolled across the two courses, and 20 consented to research use(Course A: 12 enrolled, 8 consented; Course B: 14 enrolled, 12 consented).

The intervention formed part of regular course activity in place of a stand-alone workshop concerning AI. Three elements were held constant across the two courses, namely the structure of the teaching partner prompt, the requirement that students record their exchanges with it, and the integration of partner use into graded coursework; instructors, for their part, adapted the disciplinary framing of the prompt, the assignments to which it was attached, and its placement within the semester. Partner use in Course A was attached to a sequence of five take-home assignments distributed across the term, whereas in Course B it was attached to weekly assignments and a mini project during the second half of the semester.

Research methods instruction offers a demanding setting for higher-order thinking because methodological problems seldom admit a single correct answer, and students must accordingly coordinate theoretical, methodological, and epistemic judgments as they clarify what a construct represents, evaluate whether evidence supports a claim, identify threats to inference, compare design alternatives, and defend their choices under competing constraints. Students in Course A, for instance, examined formative and reflective measurement models, the alignment among theory, hypotheses, and methods, common method bias, and trade-offs between internal and external validity. The setting thus supplies both an authentic instructional context and observable artifacts through which higher-order thinking may be studied.

### 3.2 The teaching partner

The teaching partner is not a system we built but a configuration of an LLM: a structured prompt, an instructor-supplied disciplinary problem, and a classroom protocol that guides students on how to use the prompt and when and how students consult it, which includes three pedagogical features. Constraint-first interaction requires that students articulate a tentative position and its supporting reasoning before consulting the teaching partner; guided dialogue directs the teaching partner to reply with reflective questions instead of answers; and epistemic framing instructs the partner to present itself as a statistical system rather than as an authority and to return judgment to the student wherever deference becomes apparent. Of the three, the constraint-first requirement does most of the work here. Students have to form a position before the model can offer one. This cannot be secured by the prompt only. The assignments co-designed between the team and instructors make that position a precondition. The redesigned assignment asks students to commit to an answer and its reasoning before any interaction with the teaching partner. This approach changes how the exchange begins. Therefore, instead of asking the model for an answer, students come to it with reasoning of their own.

Instructors frame the disciplinary problem within the prompt template (see Appendix) provided by the team, and students paste and adapt the prompt. In an illustrative prompt drawn from a statistics section, the partner is instructed not to supply answers or to run analyses, to require the student's own position and reasoning first, to respond only with questions which press the student to justify, test, or revise that position, and to hand judgment back should the student begin deferring; the prompt then names the disciplinary problem, in this instance whether a significant treatment effect in an observational education dataset may be interpreted causally, and offers example questions the partner might pose.

### 3.3 Design cycles

The conjecture map tested here emerged from an earlier cycle conducted in Spring 2025 within a single undergraduate research methods course (N = 14), which generated 442 student-AI messages and identified two contrasting epistemic positions, validator and co-thinker. This cycle is reported in a companion manuscript by the same authors, provided in anonymized form in the concurrent submissions field.

Phase 1 extends the design in three respects. It moves from a single undergraduate course into graduate coursework spanning two disciplinary contexts; it adds systematic capture of the interaction record as a data source in its own right, where the earlier cycle had treated such records as an instructional byproduct; and it adds measures of AI literacy and metacognitive awareness alongside critical thinking, following the earlier cycle's observation that students routinely attributed intention to the model.

### 3.4 Measures

AI literacy and higher-order thinking, including critical thinking, serve as co-primary outcomes. The measures fall into three layers supporting different kinds of claims, and they are distinguished in advance here because the distinction governs how the results should be read.

Three instruments asked students to appraise their own capabilities. AI literacy, understood as the competencies required to evaluate and reason about AI systems (Long & Magerko, 2020), was measured using the Meta AI Literacy Scale (MAILS), which comprises nine dimensions and a total score (Brown, 1992; Koch et al., 2024). Critical thinking disposition was measured with an adapted 13-item scale comprising Reasoning (8 items) and Open-Mindedness (5 items) rated on five-point agreement scales, yielding a possible range of 13 to 65. Metacognitive awareness was measured with a 19-item scale adapted from the factor structure of the Metacognitive Awareness Inventory (Schraw & Dennison, 1994), comprising Knowledge of Cognition (7 items) and Regulation of Cognition (12 items distributed across planning, monitoring, information management, debugging, and evaluation), with a possible range of 19 to 95.

The critical thinking and metacognition scales were adapted for this study, not administered in previously validated form. Internal consistency for the critical thinking scale was marginal at pretest (alpha = .64) and high at posttest (alpha = .91), while the metacognition scale was adequate to high at both administrations (alpha = .82 and .94); these values are reported here instead of alongside the results because the marginal pretest reliability bears directly on how small pre-post differences ought to be interpreted.

Course B's learning assessment included three objectively scored knowledge items assessing course concepts independently of tool use, and these constitute the only measures in the study which assess performance in place of self-appraisal with the partner absent. Finally, students' interaction with the teaching partner was captured as complete conversation transcripts; this record is the only source in the study which evidences engagement as enacted, not as reported, and it accordingly occupies a central place in the analysis.

### 3.5 Interaction corpus

Students submitted transcripts of their sessions with the teaching partner alongside their course assignments, and the metadata record comprises 104 submissions from 21 students across 95 transcript files.

Corpus construction revealed a pattern of data loss which we report as a finding concerning instrumentation rather than as a technical footnote. Nine submissions contained no retrievable content, chiefly private conversation URLs where a shareable link had been required; and of the 95 retrievable files, 20 could not be segmented into turns because they contained no speaker markup, these being pasted extracts, several of which consist of the model's response with the student's own contributions absent. All 20 originated in Course B, and all were substantially shorter than the segmentable files, averaging 3,688 characters against 22,730.

The resulting corpus is unbalanced. The analyzable record comprises 75 transcripts from 16 students and yields 594 student turns; after removing 65 opening turns consisting of the pasted system prompt, 529 analytic student turns remain, of which 408 (77%) come from Course A and 121 (23%) from Course B. Course A contributed all eight of its students to this record, whereas Course B contributed eight of thirteen.

### 3.6 Coding of student prompts

Epistemic positioning and cognitive process were coded from student turns using a scheme fixed before coding began and documented with its date. The unit of analysis is the student prompt, defined as a single message sent by a student to the AI assistant. Each prompt receives one cognitive process code drawn from the upper registers of the revised Bloom taxonomy (analyze, evaluate, create, or procedural; Anderson & Krathwohl, 2001), a metacognitive regulation flag recorded separately so that co-occurrence may be examined, and, unless procedural, one epistemic positioning code.

Positioning codes follow the three-position model developed in the earlier cycle. A turn is coded as a validator when the student treats the partner as an authority whose role is to confirm or correct work already produced, as a co-thinker when the student brings a position and engages the partner's response critically, and as a generator when the student asks the partner to produce content without first advancing a position of their own. An open category records turns fitting none of these positions, together with a descriptive phrase, so that the adequacy of the three-position model may be assessed instead of being assumed.

Coder 4, an educational researcher trained in classroom discourse analysis, worked independently to analyze the student prompts for their conversational uptake of the AI Assistant's prompts. Conversational uptake refers to the ways in which a speaker builds on the contributions of their interlocutor (i.e., AI Assistant) (Scheinberg et al., 2026). Coding began from the assumption that high uptake elaborates on the interlocutor's contributions; low uptake does not move a topic forward (Demszky et al., 2021). Using an inductive approach, Coder 4 created codes to describe uptake strategies. These included the following codes when conversational uptake was present in students' subsequent prompts: (1) *Respond*: Student responded directly to a question or suggestion from the interlocutor; (2) *Ask*: Student asked a question related to a topic raised by the interlocutor; (3) *Revise*: Student revised a question, claim, or idea based on a topic raised by the interlocutor; and (4) *Challenge*: Student challenged an aspect of the interlocutor's prompt by indicating an error, pointing out irrelevance, or stating their confusion. When conversational uptake was not present, the following codes were used: (1) *Ignore*: Interlocutor directly asked for a response, but student did not respond; and (2) *Redirect*: Student directed conversation to a new topic by asking a question, sharing a claim, or making an argument unconnected to the interlocutor's previous prompt. Illustrative examples are found in the findings.

### 3.7 Analysis

Quantitative analyses used all available matched cases for each instrument, yielding analytic samples of 18 for the MAILS, 19 for critical thinking and metacognition, 11 for Course B's learning assessment, and 20 for demographics; all scored items in the analyzed records were complete and within their allowable ranges. Primary analyses used paired-samples t tests with Cohen's dz as the paired-change effect size, and multiplicity was controlled through Holm adjustment within two outcome families, namely the MAILS family comprising nine dimensions and the total, and the critical thinking and metacognition family comprising 11 outcomes. Exact Wilcoxon signed-rank tests, bootstrap confidence intervals, leave-one-out analyses, outlier checks, and attention-check sensitivity analyses were used to assess robustness. Because the analytic samples are small, a sensitivity power analysis is reported alongside the results, and non-significant findings are interpreted as uninformative about change rather than as evidence that no change occurred.

Interaction analyses characterize the corpus descriptively and examine within-semester patterns across Course A's sequenced assignments, which constitute the study's only ordered series. Analyses of conversational uptake occurred by creating different data displays to identify larger themes to describe the data set. Specifically, we examined code application by course, assignment, and student, and code co-occurrence to identify patterns of conversational uptake strategies from high to low.

### 3.8 Ethics

This study was approved by the Institutional Review Board of the authors' university (Protocol IRB-23-07810-XP). Students consented to the research use of course-generated data; participation had no bearing on course standing, and instructional activities proceeded for all enrolled students irrespective of consent; interaction records were de-identified prior to analysis and are stored separately from any linking information.

### 3.9 Positionality

The first author designed the intervention and the study but taught neither course, and our positionality therefore concerns our relationship to the instructors whose practice we studied, not as any dual role as instructor and researcher. The instructors participated as colleagues within the same college, and the design team depended on them both for implementation and for the return of data; we recognize that this dependence has informed what was reported to us concerning implementation, and that our account of enactment consequently depends more on artifacts than on instructor self-report. The second author joined the team in Fall 2025, first as an instructor piloting the intervention in a community-based research design course for doctoral students and later as a qualitative researcher with expertise in design-based research methodology. The third author served as a teaching assistant during the initial phase of the study and helped implement the instructional cycle and worked on the project's quantitative analysis; this earlier involvement informed his interpretation of the data. The fourth author joined the project during the initial development phase. He served as a liaison team member for Course A, collaborating with its instructor on AI integration in assignment design and data collection. He has research expertise in human-AI interaction and computational social science.

# 4 Findings

## 4.1 Sample

Twenty students across the two courses contributed demographic data, 12 (60%) from Course B and 8 (40%) from Course A. The sample was predominantly graduate or professional (95%), enrolled in a college of education, health, and human sciences (80%), and women (75%); ages ranged from 21 to 49, with most participants between 25 and 39 (80%). Seven students (35%) were first-generation college students, and 11 (55%) were international students; fourteen (70%) were employed, and half reported dependents, and the modal household income bracket was under $40,000. Access to learning technology was high; all students reported regular computer access, 19 (95%) reported reliable internet, and 17 (85%) rated their digital literacy as advanced (Table 2). These characteristics describe an older, largely graduate sample possessing strong digital access alongside heterogeneous educational, family, and economic circumstances.

Table 2.

Participant characteristics

| **Characteristic** | **Category** | **n** | **%** |
|---|---|---|---|
| Course | Course A | 8 | 40.0 |
| | Course B | 12 | 60.0 |
| Primary college | Education, health, and human sciences | 16 | 80.0 |
| | Arts and sciences | 4 | 20.0 |

| Characteristic | Category | n | % |
|---|---|---|---|
| Class standing | Graduate or professional | 19 | 95.0 |
| Gender | Woman | 15 | 75.0 |
| | Man | 5 | 25.0 |
| Age (years) | 21-29 | 9 | 45.0 |
| | 30-39 | 9 | 45.0 |
| | 40-49 | 2 | 10.0 |
| Neurodiversity | Yes | 6 | 30.0 |
| First-generation college student | Yes | 7 | 35.0 |
| Employment | Full-time | 8 | 40.0 |
| | Part-time | 6 | 30.0 |
| | Not employed | 6 | 30.0 |
| International student | Yes | 11 | 55.0 |
| Has dependents | Yes | 10 | 50.0 |
| Digital literacy | Advanced | 17 | 85.0 |
| | Intermediate | 3 | 15.0 |
| Regular computer access | Yes | 20 | 100.0 |
| Reliable internet access | Yes | 19 | 95.0 |
| Household income | Less than $40,000 | 11 | 55.0 |

| Characteristic | Category | n | % |
| --- | --- | --- | --- |
| | $40,000-$99,999 | 5 | 25.0 |
| | $100,000 or more | 4 | 20.0 |

*Note.* N = 20. Percentages are based on nonmissing responses. For binary indicator rows only the affirmative category is displayed. First-generation status was derived from the absence of a parental or caregiver bachelor's degree. Age bands and income categories were collapsed, and one further category suppressed, to protect participants in small cells.

### 4.2 Self-reported capability increased

MAILS total scores increased from pretest ($M$ = 204.00, $SD$ = 41.16) to posttest ($M$ = 228.72, $SD$ = 40.32), a gain of 24.72 points, 95% $CI$ [11.58, 37.86], $t(17)$ = 3.97, Holm-adjusted $p$ = .009, $dz$ = .94; the result was consistent under an exact Wilcoxon signed-rank test ($p < .001$), a bootstrap interval for the mean gain of [13.11, 36.72], and leave-one-out analyses (all $p < .003$).

The dimension-level pattern proved narrower than the total. Understanding AI increased from $M$ = 37.56 ($SD$ = 9.41) to $M$ = 43.28 ($SD$ = 8.91), Holm-adjusted $p$ = .008, $dz$ = .96, and was the only dimension to survive correction, while AI Learning showed a nominal increase that did not ($dz$ = .53, adjusted $p$ = .302). Taken together, the pattern supports a broad increase in the total while locating the clearest dimension-specific change in students' conceptual understanding of AI rather than across the facets concerning use, creation, ethics, or self-regulation (Table 3).

Within Course B's learning assessment (n = 11), five self-rated competencies survived Holm adjustment with large effects, namely reproducible workflows ($dz$ = 1.09), working with external or API data ($dz$ = 1.56), code-based visualization ($dz$ = 1.26), geographic and spatial data ($dz$ = 1.36), and evaluating AI-generated code or explanations ($dz$ = 1.56); confidence in reading another person's code and in data cleaning showed nominal gains which did not survive correction (Table 4). Course A's pretest and posttest assessed different content and therefore cannot support a paired learning-gain analysis, the pretest having covered research methods knowledge and confidence while the posttest covered confidence with statistics and with AI-supported statistics learning.

Table 3.

Pretest-posttest changes in AI literacy, critical thinking, and metacognition

| **Outcome** | **n** | **Pre *M* (*SD*)** | **Post *M* (*SD*)** | **Delta *M*** | ***p* Holm** | ***dz*** |
|---|---|---|---|---|---|---|
| ***Panel A. Meta AI Literacy Scale*** | | | | | | |
| Apply AI | 18 | 40.61 (12.45) | 42.89 (13.45) | 2.28 | .782 | 0.35 |
| Understand AI | 18 | 37.56 (9.41) | 43.28 (8.91) | 5.72 | .008 | 0.96 |
| **Detect AI** | **18** | **21.33 (3.60)** | **22.89 (4.74)** | **1.56** | **.782** | **0.35** |
| AI ethics | 18 | 22.94 (4.61) | 24.22 (4.92) | 1.28 | .782 | 0.23 |
| Create AI | 18 | 4.83 (7.38) | 9.44 (10.37) | 4.61 | .445 | 0.45 |
| AI problem solving | 18 | 18.39 (5.39) | 21.00 (4.74) | 2.61 | .427 | 0.47 |
| AI learning | 18 | 14.33 (6.44) | 17.67 (5.26) | 3.33 | .302 | 0.53 |
| AI persuasion literacy | 18 | 22.33 (5.53) | 22.89 (5.82) | 0.56 | .782 | 0.10 |
| AI emotion regulation | 18 | 21.67 (7.55) | 24.44 (4.96) | 2.78 | .782 | 0.34 |
| Total AI literacy | 18 | 204.00 (41.16) | 228.72 (40.32) | 24.72 | .009 | 0.94 |
| ***Panel B. Critical thinking and metacognition*** | | | | | | |

| **Outcome** | **n** | **Pre *M* (*SD*)** | **Post *M* (*SD*)** | **Delta *M*** | ***p* Holm** | ***dz*** |
|---|---|---|---|---|---|---|
| Reasoning | 19 | 33.53 (2.87) | 35.42 (3.50) | 1.89 | .468 | 0.50 |
| Open-mindedness | 19 | 21.63 (1.92) | 21.68 (2.69) | 0.05 | > .999 | 0.02 |
| Critical thinking total | 19 | 55.16 (3.53) | 57.11 (5.78) | 1.95 | > .999 | 0.35 |
| Knowledge of cognition | 19 | 27.74 (3.59) | 29.26 (4.71) | 1.53 | > .999 | 0.36 |
| Planning | 19 | 8.42 (1.54) | 7.95 (1.47) | -0.47 | > .999 | -0.28 |
| Monitoring | 19 | 7.32 (1.42) | 7.05 (1.96) | -0.26 | > .999 | -0.17 |
| Information management | 19 | 8.53 (1.26) | 8.26 (1.52) | -0.26 | > .999 | -0.20 |
| Debugging | 19 | 8.42 (1.22) | 8.47 (1.31) | 0.05 | > .999 | 0.03 |
| Evaluation | 19 | 15.79 (2.12) | 15.68 (3.50) | -0.11 | > .999 | -0.04 |
| Regulation of cognition | 19 | 48.47 (5.09) | 47.42 (7.81) | -1.05 | > .999 | -0.16 |
| Metacognition total | 19 | 76.21 (7.48) | 76.68 (12.20) | 0.47 | > .999 | 0.05 |

*Note.* Delta M = posttest minus pretest; dz = Cohen's standardized paired-change effect. Holm-adjusted p-values were calculated separately within the MAILS family (nine dimensions plus the total) and the critical thinking and metacognition family (11 outcomes). Because the MAILS total is not independent of its nine constituent dimensions, including both within the same family is conservative with respect to the total. Boldface marks results remaining significant at alpha = .05 after adjustment.

Table 4.

Course B learning assessment pretest-posttest results

| Item | Pre *M* (*SD*) | Post *M* (*SD*) | Delta *M* | *p* Holm | *dz* |
|---|---|---|---|---|---|
| **Reproducible workflows (R Markdown)** | **1.73 (0.65)** | **2.64 (0.50)** | **0.91** | **.028** | **1.09** |
| Interpret another person's R code | 1.91 (0.30) | 2.45 (0.52) | 0.55 | .126 | 0.79 |
| Data cleaning and preparation | 1.91 (0.54) | 2.45 (0.52) | 0.55 | .126 | 0.79 |
| Meaning of data pivoting (objective) | 0.82 (0.40) | 1.00 (0.00) | 0.18 | .334 | 0.45 |
| **External or API data** | **1.55 (0.52)** | **2.27 (0.47)** | **0.73** | **.004** | **1.56** |
| **Code-based visualization** | **2.18 (0.40)** | **2.82 (0.40)** | **0.64** | **.013** | **1.26** |
| Purpose of faceting (objective) | 1.00 (0.00) | 0.91 (0.30) | -0.09 | .341 | -0.30 |
| **Geographic or spatial data** | **1.18 (0.40)** | **2.00 (0.45)** | **0.82** | **.009** | **1.36** |
| Meaning of tokenization (objective) | 0.64 (0.50) | 0.91 (0.30) | 0.27 | .245 | 0.58 |
| **Evaluate AI-generated code or explanations** | **1.64 (0.67)** | **2.36 (0.50)** | **0.73** | **.004** | **1.56** |

*Note.* n = 11 matched students. Self-rating items used a 1-3 scale and objective items a 0-1 incorrect/correct scale. Holm-adjusted p-values control the family of 10 item tests. Boldface marks results remain significant at alpha = .05 after adjustment.

### 4.3 Demonstrated performance did not increase

Critical thinking disposition rose modestly, from $M = 55.16$ ($SD = 3.53$) to $M = 57.11$ ($SD = 5.78$), a change of 1.95 points, 95% *CI* [-0.70, 4.59], Holm-adjusted $p > .999$, $dz = .35$; the exact Wilcoxon result was likewise nonsignificant, and the bootstrap interval included zero. Reasoning reached nominal significance (*dz* = .50) though not after correction, and Open-Mindedness was essentially unchanged. Metacognitive awareness was stable (change = 0.47, *dz* = .05), no subscale supporting reliable change; and in a sensitivity analysis restricted to the 17 students who passed the attention check at both administrations, the critical thinking total reached nominal significance (*dz* = .58) but remained nonsignificant after familywise correction. None of the three objectively scored knowledge items in Course B's assessment changed.

These null results must be read in light of the study's sensitivity. With n = 19 and *alpha* = .05, a paired t test had 80% power to detect a standardized paired change of approximately *dz* = 0.67 or larger, and the observed effects fall well below that threshold; the analysis was accordingly not capable of detecting changes of the magnitude a one-semester intervention might plausibly produce, and we therefore treat these results as uninformative about small or moderate change over evidence that no change occurred. Ceiling effects compound the limitation, since baseline critical thinking scores already stood at 81% of the usable scale range with 10.5% of students near the ceiling, and pretest reliability was marginal.

What organizes the results across Sections 4.2 and 4.3 is thus not principally a distinction between constructs proximal and distal to instruction; every instrument which registered change was a self-report, and every measure of demonstrated performance was flat.

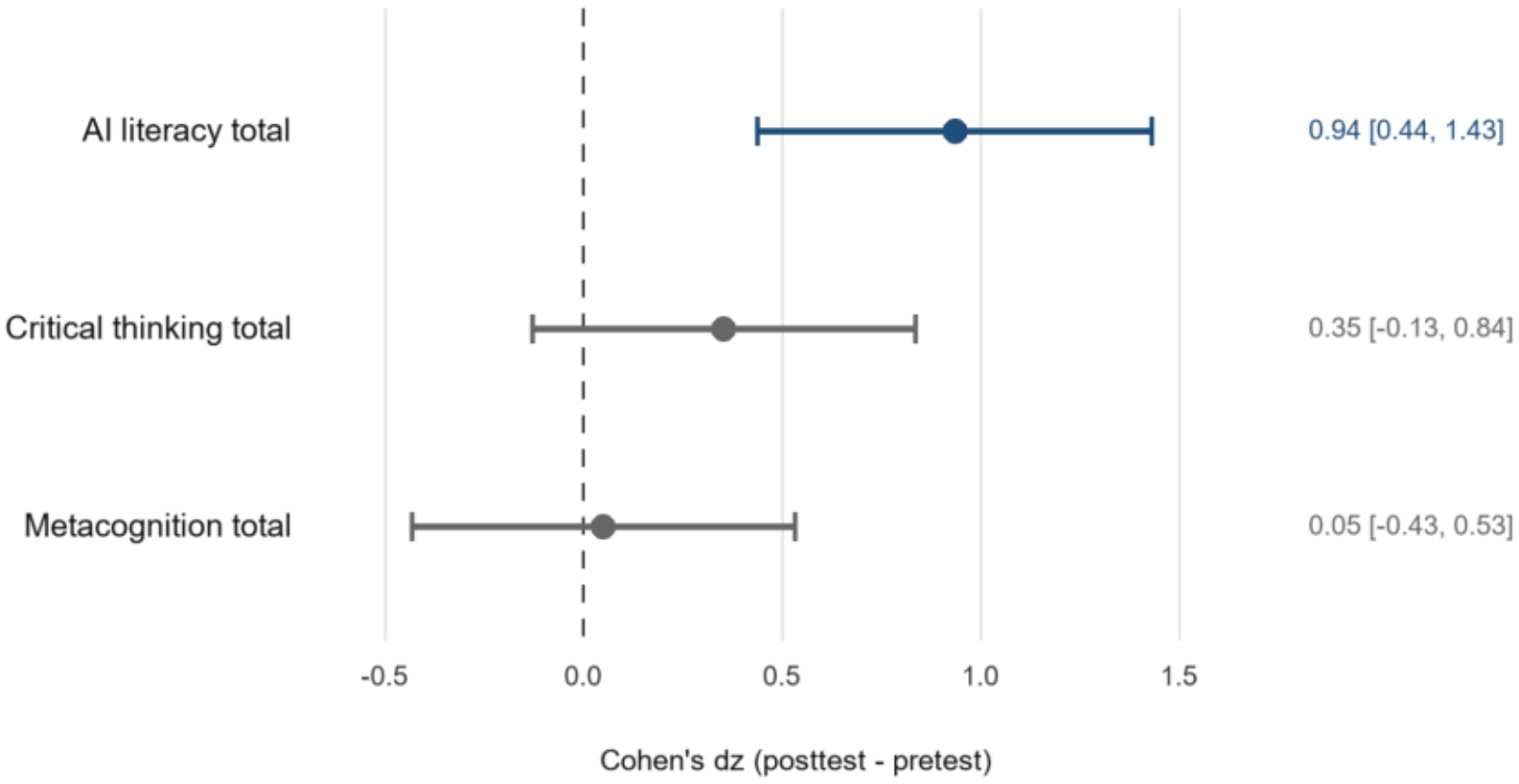

Figure 1**:** Standardized paired-change effects for the three outcome totals.

*Note.* Points are Cohen's dz and whiskers are 95% confidence intervals on the dz scale, which differ from the mean-change intervals in scale points reported in Table 3. The dashed vertical line marks no change. Positive values indicate higher posttest scores. Each series is labelled and each estimate printed, so the figure does not rely on colour to convey information.

### 4.4 The interaction record

Students wrote a median of 38.5 words per turn in Course A and 49 in Course B, against a model median of 338; exchanges were brief, with a median of five student turns per transcript, a mean of 7.6, and a range from 1 to 45, and 21.6% of student turns ran under 20 words.

Two lexical indicators, reported descriptively, bear on whether the design was enacted as intended. Forty-eight percent of student turns contained a question, with little difference between courses (47% in Course A and 51% in Course B), whereas 25.5% contained an explicit statement of the student's own position, marked by phrases such as “I think,” “my reasoning,” or “here is what I wrote,” again with little difference between courses (24.8% and 28.1%). Since the constraint-first sequence requires position-stating at the opening of an exchange rather than in every turn, this proportion understates compliance; it does nonetheless suggest that across the corpus as a whole, students asked more often than they asserted.

Course A's five sequenced assignments permit a within-semester trajectory (Table 5). Median turn length rose from 37 words on the first assignment to 44 on the last, and the proportion of turns containing a question moved from 54% to 78%, although neither trend is monotonic and the final assignment rests on six students and 45 turns; turns per student showed no trend.

Table 5.

Course A interaction across sequenced assignments

| **Assignment** | **Students** | **Turns** | **Turns per student** | **Median words** | **% with question** |
|---|---|---|---|---|---|
| 1 | 7 | 72 | 10.3 | 37 | 54.2 |
| 2 | 7 | 98 | 14.0 | 31 | 35.7 |
| 3 | 8 | 83 | 10.4 | 38 | 53.0 |
| 4 | 7 | 91 | 13.0 | 44 | 42.9 |
| 5 | 6 | 45 | 7.5 | 44 | 77.8 |

*Note.* Analytic student turns only; opening turns consisting of the pasted system prompt are excluded. Median words refer to student turns. Percentage with question denotes the proportion of student turns containing at least one question mark, a surface indicator reported descriptively instead of a coded construct.

A third feature of the record helps make sense of the metacognition result. Students explicitly described changing a position they had brought into the exchange, and two linked the change to the partner's questioning. All three passages below come from Course A, which also supplies most of the corpus. One student wrote that the partner's question was "making me uncomfortable in a useful way." The student then recognized an impulse to keep adding mediators until one worked, named the impulse as "exactly the post-hoc rationalization" the course had warned against, and concluded that a null mediation result should instead raise doubts about the proposed mechanism. A second student, on the fourth assignment of the semester, looked back at an initial analysis as "much more surface level" because it had failed to question the design, concluding that the theoretical reading needed to come first. The same student had earlier, on the first assignment, pinpointed what changed a judgment about a measurement model, namely the partner's question about whether the construct could survive if one of its dimensions were removed.

We treat these passages as illustrative, not as estimates of how often revision occurred. Our identification was lexical, so it would miss changes in reasoning that emerged across turns without being explicitly stated. Even with this limitation, these passages matter because they are the only evidence in this study that shows students noticing and revising their own reasoning as it happened. The evidence is present alongside a metacognition scale on which no subscale changed.

In analyzing conversational uptake, we found 448 turns in which the AI Assistant directly asked the student to respond (as expected under the constraint of a dialogic thought partner). In most instances when the AI did not directly seek a response, students had not begun with the constraint-first prompt. Students mostly took up questions or suggestions from the AI Assistant (i.e., *Respond*) in their subsequent prompts, while ignoring (i.e., *Ignore*) the questions or suggestions approximately 25% of the time. An example response was:

**AI Assistant:** “What would that tell you theoretically?”

**Student:** "This would prevent the chain of the model, but it would show me where the model needs improving."

When students ignored the AI Assistant's prompt, they typically asked an unrelated question to redirect the conversation. Especially for Course A, evidence suggested that these unrelated questions were provided by the instructor (i.e., identical questions were found across an assignment for multiple students). This pattern of no conversational uptake involved students entering instructor questions or prompts into the chat, without engaging the AI Assistant's responses (at least within the chat). The AI Assistant followed the students' lead.

For those who responded, the nature of students' conversational uptake varied across students, assignments, and courses. Around three-fourths of student turns containing responses came from Course A, and conversational uptake was noticeably higher for two assignments (second and fourth). Conversational uptake remained even (and low) across assignments and students for Course B. One student in Course A showed particularly high conversational uptake across all five assignments. This students' high conversational uptake suggested a pattern in which students engaged with the AI Assistant's prompts in a number of ways, including directly answering questions (i.e., *Respond*); asking their own related questions (i.e., *Ask;* for example: "Thank you for clarifying…Should I then consider demand characteristics and weak manipulation?"); sharing revised claims or arguments (i.e., *Revise*; for example: AI: "What does that tell you about your original answer?" S: "I can see now that the student was correct because they are making causal claims."); and challenging the AI Assistant (i.e., *Challenge*; for example: "Actually I am not sure if this is right because I didn't give you the data."). Around half of students' responses included one of these strategies, with questions being most common.

Two patterns related to low conversational uptake emerged. For the first pattern, students demonstrated high conversational uptake early in the dialogue, but eventually challenged the AI Assistant and/or redirected the conversation more frequently in later parts of the conversation. For example: "I understand in each section you will have questions and its never ending. However, I want to think [about] this question…[*unrelated questioned*]." The other pattern of low conversational uptake involved a brief response to the AI Assistant's question or suggestion followed immediately by an unrelated question or claim/argument to change the direction of the conversation. For example:

> **AI Assistant:** How does a "transparent and honest" researcher navigate a situation where the Math and the Human Experience disagree?
>
> **Student:** I would report both truths and make the tension explicit. Now… [*unrelated question from instructor*].

Moreover, this pattern of low conversational uptake included basic acknowledgements, such as "I understand," followed by unrelated questions or claims/arguments to redirect the conversation.

### 4.5 What the corpus indicates about enactment

The difficulties of corpus construction described in Section 3.5 constitute evidence concerning the design as implemented. Twenty of 95 submissions, all originating in one course, contained no speaker markup and several preserved only the model's output, while a further nine contained no retrievable content at all; together these account for roughly 28% of submissions.

The implication extends beyond data loss. A submission consisting of the model's response with the student's contributions absent is consistent with a workflow where the exchange was understood as a product to be handed in over a record of reasoning to be preserved; and while we cannot establish that interpretation from the artifacts alone, we can say that the design presumed a dialogic record and that in a substantial minority of cases no such record was produced.

## 5 Discussion

### 5.1 The conjecture, partially tested

The conjecture map holds that three design features shift students away from confirmation-seeking and toward reasoning, and that this shift in turn produces gains in higher-order thinking and AI literacy; the present cycle supports part of that chain while leaving its central link open.

AI literacy increased substantially, and the increase concentrated in students' conceptual understanding of AI, which is what one would expect from a semester of structured contact with a system framed explicitly as a statistical model in place of an authority. Students who spend a term being told that the system is a statistical model, and who repeatedly encounter its refusal to supply answers, appear to come away with a clearer sense of what such systems are; this is consistent with the epistemic framing feature operating as designed, though consistency is of course a weaker warrant than the conjecture map ultimately requires. The link running from positioning to higher-order thinking remains untested rather than disconfirmed, since our measures of higher-order thinking consisted of self-report dispositions together with a small set of objective items and the study lacked the sensitivity to detect anything short of a large effect.

### 5.2 A measurement problem in this literature

Our results admit two readings which the survey evidence alone cannot distinguish. On the first reading, students gained in the practices directly rehearsed during the semester while broader dispositions changed little, which would constitute an unremarkable pattern of proximal against distal transfer; on the second, a semester spent with a teaching partner reliably produced change in how students appraise their own capability, unaccompanied by any performance evidence that the underlying thinking changed.

We take this to be a difficulty for the wider literature and not as a peculiarity of the present study. The work summarized in Table 1 evaluates designs through instructor perception, learner self-report, and design exploration; and if self-appraisal and demonstrated performance can diverge as they did here, then a field relying on the former is not well positioned to know whether its designs achieve the latter. The instruments in common use, including those we ourselves selected, were not constructed to make that distinction. Naturally the implication we draw is a narrow one concerning evidentiary status instead of a general preference among methods: interaction records warrant treatment as primary evidence about mechanism, since they constitute the only source which observes engagement instead of asking about it, and a design study reporting what learners say about a tool while saying nothing about what they did with it has not tested its own conjecture.

### 5.3 Implications for design

The enacted record suggests three revisions to the design. The constraint-first sequence appears to require structural over instructional enforcement, since instructing students to state a position before consulting the model produced position-stating in roughly a quarter of turns, whereas a design which cannot proceed until the student has entered their own reasoning would enforce what the prompt at present merely requests. The interaction record itself ought to be an artifact of the system, in place of student submission behavior, since more than a quarter of submissions in this cycle proved unusable or partial and concentrated within a single course, which represents both a data problem and a signal concerning how students understood the activity; system-side capture would address both at once. Exchange length and conversational uptake, finally, warrant treatment as a design target in its own right, since a median of five student turns and low conversational uptake are thin for the sustained dialogue the conjecture depends on, and revisions which lengthen exchanges and promote higher uptake may prove more consequential than revisions to prompt wording.

These revisions have practical implications. The design has since moved into a second phase with two instructors, where the changes indicated above are already in implementation; instructors now receive explicit implementation guidance instead of adapting the prompt at their own discretion, and assignments and activities are revised in advance so that the constraint-first sequence is enacted in the task structure and not merely requested in the prompt. Survey completion is now a course requirement, and learning outcomes are assessed objectively in place of self-rating. Whether these revisions produce the effects conjectured here is the question the next cycle addresses.

### 5.4 Limitations

Several limitations bound what this study can claim. The design was implemented in four courses, of which two returned no data following departures from the collection protocol, so that all results derive from the two courses in which the protocol was followed; and because implementation quality and data return are plausibly related, the analytic sample may be selected on characteristics correlated with fidelity, and the results may not represent the design as delivered across all four sites. The corpus is moreover unbalanced, 77% of analyzable student turns originating in one course and the other contributing only eight of its thirteen students to the turn-level record, so that our interaction findings describe Course A considerably better than Course B.

Two of the three self-report instruments were adapted and were not administered in validated form; pretest reliability for critical thinking was marginal, and baseline scores on the scale were near the ceiling. Every outcome that changed was self-reported, and the objectively scored items, of which there were only three, did not change; the survey evidence alone therefore cannot distinguish growth in capability from growth in confidence. Analytic samples ranged from 11 to 20, and the study had 80% power to detect only large standardized changes, so that null results are uninformative and are not evidence of absence; the design was single-arm pre-post across one semester at one institution, no causal claims are warranted, and one semester is in any case a short window for change in dispositions which ordinarily develop across programs of study.

The question and position-statement proportions reported in Section 4.4 are surface features, not coded constructs, and understate constraint-first compliance for the reason given there. Finally, the first author designed the intervention and depended on colleagues to implement it, which may have influenced what was reported back concerning implementation.

## 6 Conclusion

The proposition that generative AI ought to be positioned as a thinking partner rather than as an answer engine is well theorized and widely designed for, yet it has seldom been subjected to empirical test. This paper specified the proposition as a conjecture and examined a first cycle in two graduate research methods courses. AI literacy increased substantially, the change concentrating in conceptual understanding of AI, while critical thinking disposition, metacognitive awareness, and objectively scored knowledge did not move; and what organizes this pattern is not principally a matter of proximity to instruction, since self-appraisal moved and demonstrated performance did not. Thus we see a design literature whose central claims concern reasoning, evaluated almost entirely through instruments which ask learners how their reasoning feels to them. Design research in this area will require interaction evidence over self-report if it is to know whether its conjectures hold, and the corpus reported here suggests that gathering such evidence is more difficult than it appears, since the mechanism of collection is itself a design problem.

# APPENDICES

## A1 Prompt Template

"You are an AI teaching assistant for a university-level course. Your primary role is to facilitate student learning by cultivating critical thinking and encouraging deep engagement with the material. Do not provide direct answers, you should prompt students to reflect on their current understanding, analyze data, and explore different perspectives.

I am a student of [insert course name]. My cultural and academic background consist of [insert relevant information]. I prefer that you speak to me as a dialogic partner.

Your responses should:

1. Encourage Reflection: Ask open-ended questions that prompt me to consider my reasoning, such as:

"How do you interpret the given data?”

“What factors may influence this outcome?”

2. Facilitate Critical Engagement: Help me consider multiple viewpoints by posing thought-provoking questions, for example:

“What alternative statistical approaches could be used here?”

“How might different assumptions change your conclusions?”

3. Support Inquiry-Based Learning: Guide me to explore course materials and credible sources while avoiding direct solutions.

4. Promote Collaborative Exploration: Encourage me to discuss my ideas with my peers and compare different approaches.

Maintain a supportive and encouraging tone, emphasizing that learning is an iterative process of exploration and revision."